\documentclass[aps,prb,superscriptaddress]{revtex4}

\usepackage{graphicx}
\usepackage{amsfonts}
\usepackage{amssymb}
\usepackage[]{amsmath}
\usepackage[]{latexsym}
\usepackage[]{float}

\usepackage{bm}

\begin{document}

\title{Chiral Symmetry and Fermion Doubling in the Zero-mode Landau Levels of \\Massless Dirac Fermions with Disorder}


\author{Tohru Kawarabayashi}
\affiliation{Department of Physics, Toho University, Funabashi 274-8510, Japan}

\author{Takahiro Honda}
\affiliation{Department of Physics, Toho University, Funabashi 274-8510, Japan}

\author{Hideo Aoki}\affiliation{Department of Physics, University of Tokyo, Hongo, Tokyo 113-0033, Japan}

\author{Yasuhiro Hatsugai}\affiliation{Institute of Physics, University of Tsukuba, Tsukuba 305-8571 Japan}

\begin{abstract}
The effect of disorder on the Landau levels of massless Dirac fermions is examined for the cases 
with and without the fermion doubling.   
To tune the doubling a tight-binding model having a complex transfer integral is 
adopted to shift the energies of  two Dirac cones, 
which is theoretically proposed earlier and realizable in cold atoms in an optical lattice.
In the absence of the fermion doubling, the $n=0$ Landau level
is shown to exhibit an anomalous sharpness even if  the disorder is uncorrelated in space (i.e., large K-K' scattering). This anomaly 
occurs when the disorder respects the chiral symmetry of the Dirac cone.
\end{abstract}

\maketitle


\section{Introduction}
Kicked off by the experimental observation of the graphene quantum Hall effect \cite{Geim,Kim}, fascination 
with massless Dirac fermions is mounting, where they appear not only in graphene but more generically in 
various systems such as 
organic metals \cite{KKS,GFMP,TSKNK},  cold atom systems in optical lattices \cite{TGUJE} and molecular 
graphene \cite{GMKGM}.  
Among these systems, the number of massless Dirac fermions is always even for solid state materials, 
which is called the ``fermion doubling" \cite{NN,Hatsugai}. 
On the other hand, manipulation of Dirac cones into single cones has been theoretically considered \cite{WHA}, 
which may be realizable in optical lattices where the Hall conductivity is detectable experimentally \cite{Mei}. 
 
Inspired by these, we explore here the effect of disorder for the Landau levels of massless Dirac fermions, in particular 
in the absence of the fermion doubling.  Specifically, the effect on the $n=0$ Landau level, which is essential to the 
anomalous Hall effect of massless Dirac fermions, is examined from the viewpoint of the fermion doubling and the symmetry 
of the system.   In our previous work \cite{KHA}, we have shown that the $n=0$ Landau level of the honeycomb 
lattice (graphene) becomes anomalously sharp even in the presence of disorder, if the disorder respects the chiral 
symmetry and is spatially correlated over a distance exceeding a few lattice constants.  Conversely, we have also 
pointed out that when the disorder is spatially uncorrelated, the $n=0$ Landau level is broadened just like the other 
Landau levels, even if the chiral symmetry is respected by the disorder. 
 
To explore the single-to-double Dirac cone crossover, here we consider a two-dimensional lattice model having two 
Dirac cones which are shifted in energy with each other as in the model proposed by Watanabe et al. \cite{WHA}. 
We examine numerically the effect of disorder with this lattice model, and have found that, even if the disorder is 
uncorrelated in space, the $n=0$ Landau levels start to become anomalously sharp as the two Dirac cones are 
energetically shifted.  Notably, the anomalous sharpness in the absence of the fermion doubling occurs when 
the disorder respects the chiral symmetry for each cone. In fact, a potential disorder, which breaks the chiral symmetry, 
washes out the sharpness. 

\section{Model and Numerical Results}
We adopt here a two-dimensional square lattice with the 
following nearest-neighbor (NN) $t$ and the next nearest-neighbor (NNN) $t'$ transfer integrals,
\begin{eqnarray}
 H & = & \sum_{\bm{r}}-tc_{\bm{r}+\bm{e}_x}^\dagger c_{\bm{r}} +(-1)^{n_x+n_y}tc_{\bm{r}+\bm{e}_y}^\dagger c_{\bm{r}} 
 \nonumber \\
  & & +it' ( c_{\bm{r}+\bm{e}_x+\bm{e}_y}^\dagger c_{\bm{r}} )+ {\rm H.c.}, \quad t,t' \in \bm{R}
\end{eqnarray}
where $\bm{r}= (n_x,n_y)$ denotes the lattice points and $\bm{e}_x=(1,0)(\bm{e}_y=(0,1))$ the unit vector in $x(y)$ direction with 
all lengths measured in units of the lattice constant. 
To realize shifted Dirac cones some transfer energies have to be 
complex, and here the NNN transfer is pure imaginary. 
Although complex transfer integrals may seem unrealistic, they 
can be realized in cold atoms in optical lattices \cite{Mei,Goldman}.
In the absence of a magnetic field, 
the Hamiltonian in the momentum space
is expressed as 
\begin{equation}
H(\bm{k})= \left[\begin{array}{cc}
 2t'\sin k_2&\Delta (\bm{k})\\
\Delta ^* (\bm{k}) & 2t'\sin k_2
\end{array}\right] 
\end{equation}
where
$\Delta (\bm{k}) = -t(-1+e^{ik_1}+e^{ik_1+ik_2}+e^{ik_2})$ with
$k_1= \bm{k}\cdot \bm{e}_1$ and $k_2 =
 \bm{k}\cdot \bm{e}_2$  
with the primitive vectors taken to be $\bm{e}_1 = \bm{e}_{x}-\bm{e}_{y}$ and 
$\bm{e}_2 = \bm{e}_{x}+\bm{e}_{y}$. In this model, we have two Dirac cones at $(k_1,k_2) = \pm (\pi/2,-\pi/2)$ with 
energies $\pm 2t'$, so that the two Dirac cones are shifted in energy 
from each other when the strength $t'$ is non-zero. A magnetic field is introduced  by taking 
the Peierls substitution $t(t') \to t(t')e^{-2\pi i \theta(\bm{r})}$, where 
the summation of the phases $\theta(\bm{r})$ along a closed loop is equal to 
the magnetic flux enclosed by it in units of $h/e$ .  
We introduce a random component $\delta t(\bm{r})$ for the NN transfer integral $t(\bm{r}) = t+\delta t(\bm{r})$ 
that has a gaussian distribution with  variance $\sigma$ and is {\it uncorrelated} in space $\langle \delta t(\bm{r}) \delta t(\bm{r}')
\rangle=\sigma^2 \delta(\bm{r}-\bm{r}')$. Still, this disorder preserves the chiral symmetry for each Dirac cone. 
The bond disorder considered in our previous work \cite{KHA} is equivalent to the present disorder in the
limit of zero correlation length.

\begin{figure}
  \includegraphics[width=0.48\textwidth]{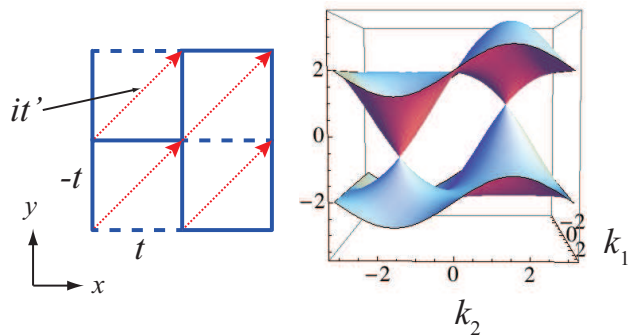}
  \caption{Left: The present lattice model. Right: Energy dispersion in the $k_1$-$k_2$ plane
   for $t' = 0.4t$. }
   \label{fig1}
\end{figure}

\begin{figure}
  \vspace*{1cm}
  \includegraphics[width=0.48\textwidth]{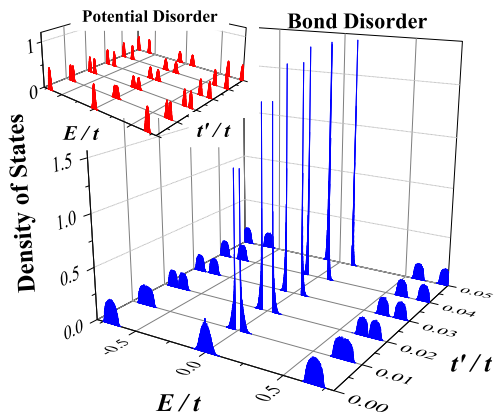}
  \caption{Density of states as function of the Fermi energy $E$ for various values of 
  $t' (\propto$ the energy shift of two Dirac cones) averaged over 
  5000 samples with a system size 30 by 30. The magnetic flux piercing each  square plaquette is $0.01(h/e)$ 
  and the strength of disorder $\sigma /t = 0.1$.
  Inset:  Density of states for the same parameters when 
  we replace the bond disorder with a potential disorder with a variance $0.1 t$. }
   \label{fig2}
\end{figure}

The numerically evaluated density of states in a magnetic field is shown in Fig.\ref{fig2}, where 
both $n=0$ and $n=\pm1$ Landau levels are split as  the 
two Dirac cones are energetically shifted with the increase of $t'$. 
Remarkably, the $n=0$ (zero-mode) Landau level for each cone becomes anomalously sharper with $t'$ even in the presence of disorder, 
while the $n=\pm1$ Landau levels remain broadened.  
For comparison, the density of states calculated for a potential disorder, which breaks the chiral symmetry, 
shows no such anomaly (Fig.\ref{fig2}, inset). 

\section{Summary and Discussions}
We have numerically investigated the effect of disorder on the Landau levels of  
the massless Dirac fermions with and without the fermion doubling. 
We have clearly shown that, 
if the chiral symmetry for Dirac cones is respected, the zero energy ($n=0$) Landau level becomes anomalously sharp 
in the absence of the fermion doubling even when the disorder is uncorrelated in space.
Since the shift in energy of the Dirac cones suppresses the mixing between the Dirac points due to the 
disorder scattering, the present result implies that the broadening reported for 
uncorrelated disorder in the honeycomb lattice (graphene) \cite{KHA} is due to the mixing 
between the two Dirac fermions with opposite chirality.


\section*{Acknowledgments}
This work is partly  supported by the Grant-in-Aid for Scientific Research Nos. 22540336 and 23340112 from JSPS.


\begin{thebibliography}{99}

\bibitem{Geim}
 K.S. Novoselov et al, 
\emph{Nature} \textbf{438}, 197 (2005).

\bibitem{Kim} Y. Zhang, Y.W. Tan, H.L. Stormer, and P. Kim, 
\emph{Nature} \textbf{438}, 201 (2005).
\bibitem{KKS} S. Katayama, A. Kobayashi, and Y. Suzuura, J. Phys. Soc. Jpn. {\bf 75}, 054705 (2006).
\bibitem{GFMP} M.O. Goerbig, J.-N. Fuchs, G. Montambeax, and F. Pi\'{e}chon, Phys. Rev. B {\bf 78}, 045415 (2008).
\bibitem{TSKNK} N. Tajima, S. Sugawara, R. Kato, Y. Nishio, and K. Kajita, Phys. Rev. Lett. {\bf 102}, 176403 (2009).
\bibitem{TGUJE} L. Tarruell et al, \emph{Nature} \textbf{483}, 302 (2012).
\bibitem{GMKGM} K.K. Gomes et al, \emph{Nature} \textbf{483}, 306 (2012). 

\bibitem{NN}
H.B. Nielsen and M. Ninomiya, \emph{Nucl. Phys. B}\textbf{185}, 20 (1981).

\bibitem{Hatsugai}
Y. Hatsugai, T. Fukui, and H. Aoki, \emph{Phys. Rev. B}\textbf{74}, 205414 (2006); Y. Hatsugai, 
\emph{J. Phys. Conf. Series} \textbf{334}, 012004 (2011).


\bibitem{WHA}
H. Watanabe, Y. Hatsugai, and H. Aoki, \emph{Phys. Rev. B}\textbf{82}, 241403(R) (2010).

\bibitem{Mei}
F. Mei et al. \emph{Phys. Rev. A}\textbf{84}, 023622 (2011).


\bibitem{KHA} 
T. Kawarabayashi, Y. Hatsugai, and H. Aoki, \emph{Phys. Rev. Lett.} \textbf{103}, 156804 (2009); \emph{Physica E}\textbf{42}, 759 (2010).


\bibitem{Goldman}
N. Goldman et al. \emph{Phys. Rev. Lett.} \textbf{103}, 035301 (2009).

\end{thebibliography}
\end{document}